\documentclass[%
 reprint,
 amsmath,amssymb,
 aps,
]{revtex4-1}

\usepackage{graphicx}
\usepackage{dcolumn}
\usepackage{bm}

\begin{document}

\title{The environment-dependence of the growth of the most massive objects in the Universe}

\author{Krzysztof Bolejko$^{1,2}$}
 \email{krzysztof.bolejko@utas.edu.au}
\affiliation{$^1$School of Natural Sciences, College of Sciences and Engineering, University of Tasmania, Private Bag 37, Hobart TAS 7001\\
$^2$Sydney Institute for Astronomy, School of Physics, A28, The University of Sydney, NSW, 2006, Australia
}

\author{Jan J. Ostrowski$^{3,4}$}
\affiliation{$^{3}$National Centre for Nuclear Research, 00-681 Warszawa, Poland \\
$^4$Univ. Lyon, Ens de Lyon, Univ. Lyon1, CNRS, Centre de Recherche Astrophysique de Lyon UMR 5574, 69007, Lyon, France 
 }%

\date{\today}

\begin{abstract}
This paper investigates the growth of the most massive cosmological objects.
We utilize the Simsilun simulation, which is based on the approximation of the silent universe. In the limit of spatial homogeneity and isotropy the silent universes reduce to the standard FLRW models. 
We show that within the approximation of the silent universe the formation of the most massive cosmological objects differs from the standard background-dependent approaches.
For objects with masses above $10^{15} M_\odot$ the effect of spatial curvature (overdense regions are characterized by positive spatial curvature) leads to measurable effects. The effect is analogous to the effect that the background cosmological model has on the formation of these objects (i.e. the higher matter density and spatial curvature the faster the growth of cosmic structures). 
We measure this by the means of the mass function and show that the mass function obtained from the Simsilun simulation has a higher amplitude at the high-mass end compared to standard mass function such as the Press-Schechter or the Tinker mass function.
For comparison, we find that the expected mass of most massive objects using  the Tinker mass function is $4.4^{+0.8}_{-0.6} \, \times 10^{15} M_\odot$, whereas for the Simsilun simulation is $6.3^{+1.0}_{-0.8} \, \times 10^{15} M_\odot$.
We conclude that the nonlinear relativistic effects could affect the formation of the most massive cosmological objects, leading to a relativistic environment-dependence of the growth rate of the most massive clusters.
\end{abstract}


\maketitle

\section{Introduction}

Observations of the most massive cosmological objects can be used as a probe for cosmology. Most often, the number of these objects and their masses is used to constrain the properties of the dark sector or the Gaussianity of primordial fluctuations.
In this paper we investigate a 
relativistic environment-dependence of the growth rate of the most massive clusters.
It is already known that growth of structure depends on the background cosmological model --- cosmological models with higher matter density and positive spatial curvature exhibit a much faster growth rate of cosmic structure than models with low matter density and negative spatial curvature.
Here we investigate whether a sufficiently large cosmic region
with positive spatial curvature (necessary for the overdensity to reach a turnaround) could exhibit a faster growth (compared to the $\Lambda$CDM model) in a similar fashion as the positively curved (globally) cosmological model.
For this purpose we utilize the framework of silent universe
\cite{1993PhRvD..47.1311M,1994PhRvL..72..320M,1995ApJ...445..958B,1997CQGra..14.1151V} and we use the Simsilun simulation \cite{2018CQGra..35b4003B}. One of the advantages of our approach is that, unlike in the perturbation schemes or most of the $N$-body simulations, the cosmological background enters our calculations only at the level of initial conditions. This allows us to trace the evolution of the gravitational instability in the far non-linear regime without setting any background by hand.

The most common approach to trace the non-linear evolution is to use the $N$-body simulations and to predict number of collapsed objects within a given mass range by constructing the mass function (eg.  \cite{2008ApJ...688..709T}).
However, standard $N$-body simulations do not include an effect referred to as 'non-kinematic differential cosmic expansion' \cite{2016JCAP...06..035B}. There is still an ongoing debate on the relevance of this effect. On one hand there are arguments in favour, suggesting that the phenomenon has real physical interpretation \cite{2013PhRvD..88h3529W,2015MNRAS.448.1660R,2016MNRAS.456L..45R,2016MNRAS.457.3285M}. 
On the other hand, there are arguments that this is merely a gauge-artifact of the comoving coordinates and can be removed by transforming to a Poisson gauge (e.g. \cite{2017arXiv171202967B}).
While such transformations and comparison are easily done within the linear approximation, within the non-linear regime, their applicability is often questioned (see e.g. \cite{2010AIPC.1241.1074M}).

In this paper we take a more pragmatic approach. We take a model of a universe (Simsilun simulation) that exhibits non-negligible backreaction effects in the comoving coordinates. We then focus on the observable quantity, in our case the mass of cosmic structures. We derive predictions as to expected number of most massive cosmic objects. 
The structure of this paper is as follows: Sec. \ref{methods} describes the methods, including the calculations of the mass functions within the Simsilun simulations; Sec. \ref{predictions} presents the predictions for the most massive objects both at high and low redshifts; Sec. \ref{conclusions} concludes the results and discusses the possibility of using the most massive cosmic objects to test and investigate the environment-dependence of the growth rate of the most massive clusters.

\section{Methods}\label{methods}

\subsection{Nonlinear relativistic evolution -- the silent universe approach}

Assuming that the source of the gravitational field is pressureless and non-rotating dust, and in addition neglecting heat and magnetic part of the Weyl tensor, then the Einstein evolution equations within 1+3 split can be reduced only to 4 scalar equations for: density $\rho$,  expansion rate $\Theta$, shear $\Sigma$), and the Weyl curvature ${\cal W}$ \cite{1971grc..conf..104E,2009GReGr..41..581E}. The evolution equations are \cite{1995ApJ...445..958B,1997CQGra..14.1151V}

\begin{eqnarray}
&& \dot \rho = -\rho\,\Theta, \label{rhot}\\
&& \dot \Theta = -\frac{1}{3}\Theta^2-\frac{1}{2}\,\kappa \rho-6\,\Sigma^2 + \Lambda,\label{thtt}\\
&& \dot \Sigma = -\frac{2}{3}\Theta\,\Sigma+\Sigma^2-{\cal W},\label{shrt}\\
&& \dot{ {\cal W}} = -\Theta\, {\cal W} -\frac{1}{2}\,\kappa \rho\,\Sigma-3\Sigma\,{\cal W},\label{weyt} 
\end{eqnarray}
where $\kappa = 8 \pi G/c^4$. Apart from these equations, one also need to satisfy the spatial constraints. However, 
the spatial constrains need only be satisfied at the initial instant, as they are conserved by the above evolution equations \cite{1997CQGra..14.1151V,1997PhRvD..55.5219M}.
Setting up the initial condition is thus important and non-trivial task. Here we follow the procedure of setting up the initial conditions in the early universe when the assumption of the Einstein--de Sitter evolution and linear perturbation is expected to work well.

Within the linear approximations, i.e. $\rho \to \bar{\rho} (1+\delta)$, where $\delta$ is the density contrast, the solution of (\ref{shrt}) and  (\ref{weyt})
are 
\[ {\cal W} = \alpha \, \kappa \bar{\rho} \, \delta \quad {\rm and} \quad \Sigma = -\frac{2}{3} \, \alpha \, \bar{\Theta} \, \delta, \]
where $\alpha$ is an arbitrary constant. 

The spatially homogeneous and isotropic FLRW models are conformally flat and shear
free, and in this case $\alpha = 0$. Setting up $\alpha = 0$ reduces the above equations to the FLRW evolutions, where different values of $\delta$ lead to different FLRW models with different parameter $\Omega$ (where $\Omega = \rho/\bar{\rho}$).
For the exact silent models such as the Lema\^itre--Tolman or Szekeres models,
the Weyl curvature is ${\cal W} = - \kappa (\rho - \rho_q)/6$, where
$\rho_q$ is a quasi-local average and $\kappa \rho_q = 6M/R^3$ \cite{2012CQGra..29f5018S,2018CQGra..35b4003B}.
If $\rho_q$ were equal to $\bar{\rho}$ then the constant $\alpha$ would be $\alpha = \alpha_q = - 1/6$.
For a sufficiently large domain and negligible spatial curvature, this indeed can be the case.
However for the formation of local overdensities, i.e. locally in the vicinity of density peaks this may no longer hold.
In the  vicinity of density peaks, a more accurate approximation, 
as verified by direct numerical calculations is $\alpha = (1/3) \, \alpha_q = -1/18$. Thus, the initial conditions for our simulations are
\begin{eqnarray}
&& \rho_i = \bar{\rho} + \Delta \rho = \bar{\rho} \, ( 1 + \delta_i), \label{rhoi} \\
&& \Theta_i = \bar{\Theta} + \Delta \Theta =  \bar{\Theta} \, ( 1 - \frac{1}{3}  \, \delta_i), \label{thti} \\
&& \Sigma_i =   \frac{1}{27} \, \bar{\Theta} \, \delta_i, \label{shri}  \\
&& {\cal W}_i = - \frac{1}{18} \, \kappa \bar{\rho} \, \delta_i \label{weyi} 
\end{eqnarray}
where  $\bar{\rho}$ and $\bar{\Theta}$ are the background density and expansion rate, and $\delta_i$ is the initial density contrast.

\subsection{The mass function}

\begin{figure}
\begin{center}
\includegraphics[scale=0.65]{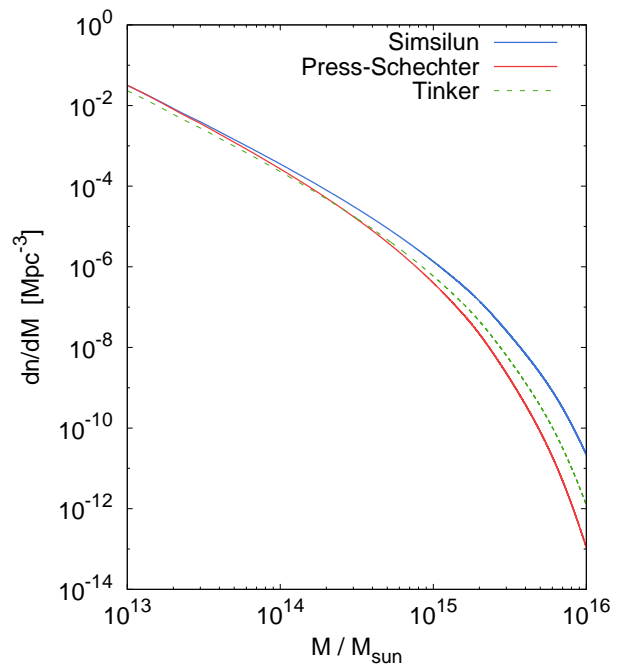}
\end{center}
\caption{Mass function at the present-day instant: the upper blue line shows the mass function evaluated within the Simsilun simulation, and the lower red line the Press-Schechter mass function; the middle dashed line presents the Tinker mass function.}
\label{fig1}
\end{figure}

\begin{figure}
\begin{center}
\includegraphics[scale=0.65]{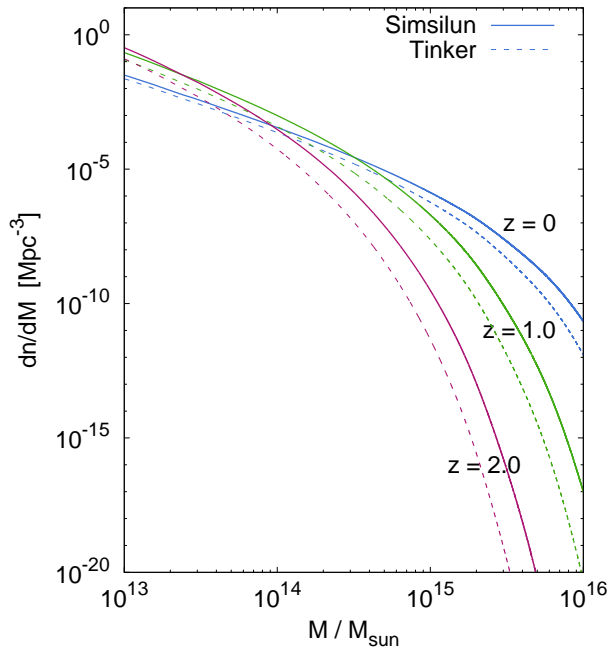}
\end{center}
\caption{Evolution of the mass function at z =0, 1, and 2. The solid line shows the mass functions evaluated using the Simsilun simulation, and for comparison the dashed line shows the Tinker mass function.}
\label{fig2}
\end{figure}

The expected number of object at a given redshift and in a given mass range can be inferred from the mass function

\begin{equation}
N = \int\limits_{z_{ {\rm min} }}^{z_{ {\rm max} }} {\rm d} z \,  \frac{ {\rm d} V} { {\rm d} z} \, 
\int\limits_{M_{ {\rm min} }}^{M_{ {\rm max} }}  {\rm d} M \frac{ {\rm d} n } { {\rm d} M},
\label{nzm}
\end{equation}
where $V$ is the volume;
and $n$ is the mass function, which  is often written in terms of the multiplicity function $f$
\begin{equation}
n(M) = f(\sigma_M) \, \frac{\rho_M}{M} \frac{ \partial \ln \sigma^{-1}_M} {\partial M},
\end{equation}
where $\sigma_M$ is the variance of the density field smoothed at scale $M$, $\rho_M$ is matter density, and $M$ is the mass.

In a generic case the multiplicity function $f$ is not an analytic function. 
There are only very limited cases where the functions $f$
has an analytic form, such as in the approach proposed by \cite{1974ApJ...187..425P}.
The Press--Schechter approach assumes density field to have Gaussian distribution and in addition
assumes that an object of mass $M$ is collapsed if the present-day linear density contrast $\delta_l$ is larger than a fixed threshold $\delta_{c}$.
In general case, the non-linear growth breaks the Gaussianity,
ie. the present-day distribution of density contrasts is no longer Gaussian but instead is much better approximated with the log-normal distribution \cite{2004LRR.....7....8L}.
Similarly, a fixed (and independent of environment) threshold $\delta_{c}$ is also a crude approximation.
Still an analytic form is useful, and in the literature one can 
find number of different parametrisations.
The discrepancy between different fits are of order of 10-20\% for masses up to $10^{15} M_\odot$, and then up to a factor of 2 for masses between $10^{15} M_\odot$ and $10^{16} M_\odot$ \cite{2013MNRAS.433.1230W,2013A&C.....3...23M}. As it is shown in Sec. \ref{masfsim}, such a level of differences between various parametrisations
is small compared to the deviation between these fits and the mass function obtained from the Simsilun simulation. Thus, in this paper we only consider one fit, i.e. the mass function of \cite{2008ApJ...688..709T}

\begin{equation}
f(\sigma) = A \, \exp \left( - \frac{c}{\sigma_M^2} \right) 
\, \left[ \left( \frac{b}{ \sigma_M} \right)^a + B \right],
\end{equation}
where 

\begin{eqnarray}
&& A = \left( 0.1 \log \Delta - 0.05 \right) \, (1+z)^{-0.14}, \nonumber \\ 
&& a = \left( 1.43 + (\log \Delta - 2.3)^{1.5} \right) \, (1+z)^{-0.06}, \nonumber \\ 
&& b = \left( 1.0 + (\log \Delta-1.6)^{-1.5}  \right) \, (1+z)^\alpha,  \nonumber \\ 
&& \log \alpha = -\left[ \frac{0.75}{ \log(\Delta/75)} \right]^{1.2},  \nonumber \\ 
&& B = 1.0, \nonumber \\ 
&& c =  1.2 (\log \Delta -2.35)^{1.6}.
\end{eqnarray}
For high density thresholds $\Delta$, such as for $\Delta > 1600$ the parameter $A = 0.26 \, (1+z)^{-0.14}$.
For comparison, the Press--Schechter function is recovered when $A =  \sqrt{2/\pi} $, $a = 1$, $b = \delta_c$, $c = { \delta_c^2}/{2}$, and $B=0$.

\subsection{The mass function of the Simsilun Simulation}\label{masfsim}

The measurements of the mass function is a useful tool that gives insight into 
cosmological properties of our universe, such as the Gaussianity of initial conditions,
or the growth of cosmic structures that may be sensitive to the properties of the dark sector or departures from the standard Einsteinian gravity. Thus in order to take a full advantage of this method we need to fully understand and appreciate the non-linear evolution of cosmic structures. 
Here, we infer the mass function using a similar approach presented by \cite{1974ApJ...187..425P}. We  first generate the initial density field with the variance given by
\begin{equation}
 \sigma_{M}^2 = \frac{1}{2\pi^2} \int\limits_0^\infty {\rm d} k\,
k^2 P(k) W(kR), \label{sigmaM}
\end{equation}
where $R$ is the radius of the object of mass $M$, i.e.  $M = (4/3) \pi R^3 \rho_M$; the  window function $W$ is
 $W(kR) = 3 \, ( \sin kR - kR \cos k R) \, (kR)^{-3}$, and $P(k)$ is the matter power spectrum $ P(k) = T(k)^2 D(z)^2 P_i$, where $P_i$ is the primordial power spectrum. The function $D(z)$ is the growth factor. The function $T(k)$ is the transfer function and is evaluated using parametrization by \cite{1998ApJ...496..605E}.

The cosmological parameters used to evaluate the initial conditions are based on the \cite{2016A&A...594A..13P}, and read $h = 0.6781$, $\Omega_b h^2 = 0.02226$, $\Omega_c h^2 = 0.1186$, $\Omega_\Lambda = 0.694$, $n = 0.9681$, and $\sigma_8 = 0.815$.

\begin{figure}
\begin{center}
\includegraphics[scale=0.7]{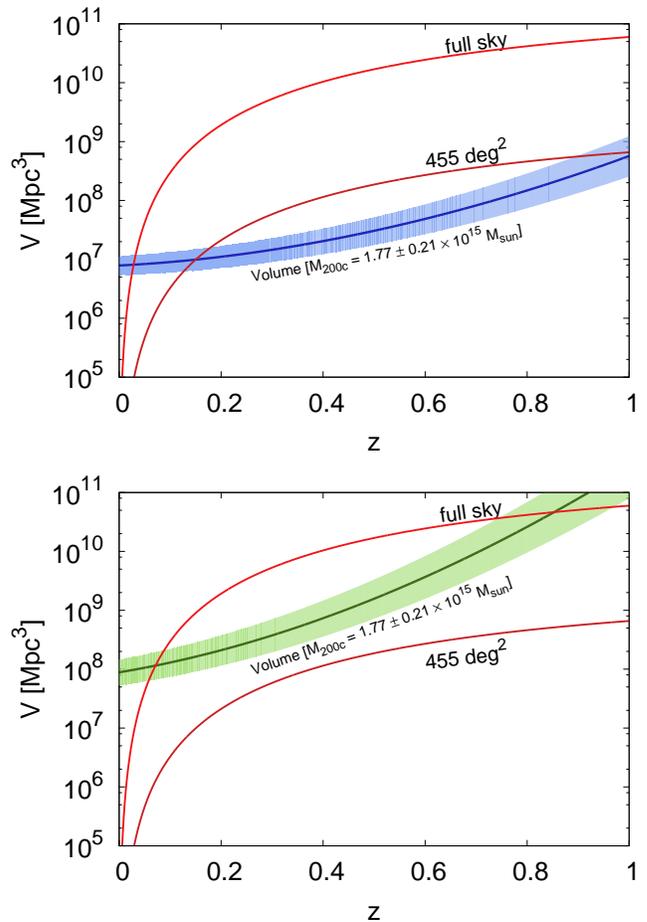}
\end{center}
\caption{Volume covered be the full-sky and ACT-like surveys (solid red lines).
For comparison the  volumes needed to observe at least 1 object evaluated using the Simsilun (upper panel) and Tinker (lower panel) mass functions are also presented. The shaded region corresponds to the mass range of $1.77 \pm 0.21 \times 10^{15} M_{\odot}$ (the density threshold $\Delta$ is set to 200 with respect to the critical density, i.e. $M_{200c}$).}
\label{fig3}
\end{figure}

\begin{figure}
\begin{center}
\includegraphics[scale=0.7]{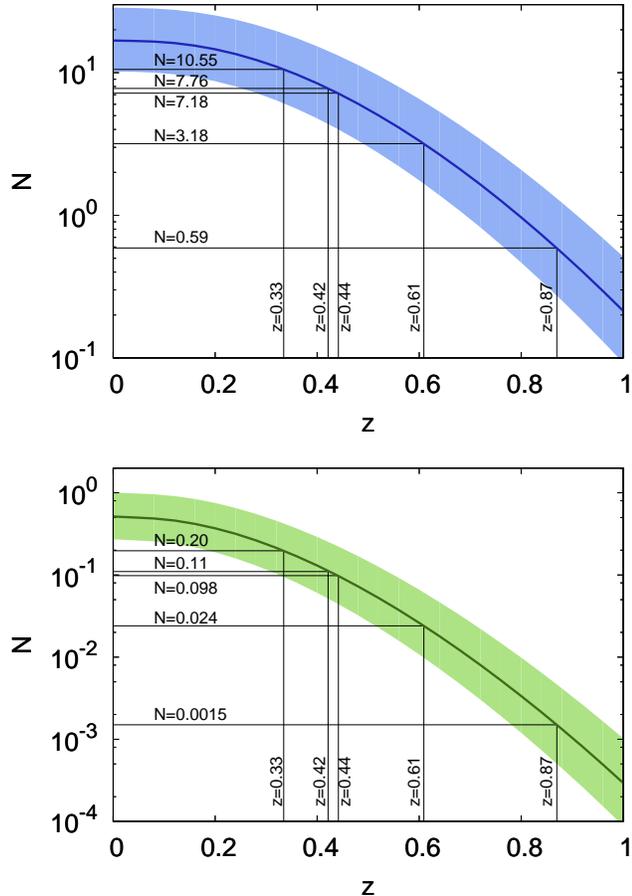}
\end{center}
\caption{Expected number of object observed within an ACT-like survey (455 deg$^2$). The number corresponds to an expected number of objects with redshift at least $z$ and masses above $1.77 \pm 0.21 \times 10^{15} M_{\odot}$.
Upper panel presents the expected number evaluated using the Simsilun mass function and the lower panel shows the expected number inferred from the Tinker mass function.}
\label{fig4}
\end{figure}

We then evolve the set of initial conditions (smoothed on different scales)  using the silent equations
(\ref{rhot})--(\ref{weyt}). 
Then, at a given time, we count the ratio
of collapsed objects to the total number of numerically generated domains to obtain the cumulative distribution and consequently the mass function. 
The collapse condition $\delta > \Delta$ can vary (where $\delta$ is the density contrast with respect to the background density, and $\Delta$ is the density contrast of the collapsed object). This condition can be fixed freely,
for example for Figs. \ref{fig1} and \ref{fig2} we use $\Delta \approx \infty$ (i.e. we stop just before the collapse is reached); this choice has been made to have a meaningful comparison with the Press-Schechter approach. However, for Figs. \ref{fig3} and \ref{fig4} we use $\Delta = 200$ with respect to the background critical density; this was done in order  to have a  meaningful comparison with the measurements obtained by the Atacama Cosmology Telescope. For Fig. \ref{fig5} we use $\Delta = 200$ but with respect to the background matter density, which has been done in order to have a direct comparison with the results obtained by \cite{2012ApJ...755L..36H} (cf. their Fig. 1).
Once the threshold is reached we stop the evolution. If we were to continue evolving pass the threshold then the region would have collapsed to a point. A typical timescale for $\Delta = 200$, from passing the threshold to reaching the final stage of the collapse is approximately 100 Myr.

Figure \ref{fig1} shows the mass function obtained as described above. 
The mass function is evaluated at the present-day instant (i.e. $z = 0$).
For comparison, the Tinker mass function is also presented.
The procedure of evaluating the mass function is similar to the procedure implemented by 
Press and Schechter, where the evolution of overdensities and their collapse is modeled with homogeneous top-hat model. In fact, if we set $\Sigma =0$, ${\cal W} = 0$ (which is equivalent to the FLRW evolution), and $\Delta=1.69$ we recover the Press-Schechter mass function, which is also presented in Fig. \ref{fig1}.

The Press--Schechter procedure does not allow for halos to merge. This is also the case of the Simsilun simulation which assumes that the worldlines do not cross (i.e. the approximation of the silent universes).
Thus, unlike in the real universe, or as observed within the $N$-body simulations,
small halos do not merge, which means that this procedure overestimates the number of objects at lower end of the mass function. There are also other effects that are relevant when comparing 
the predictions from simulations with observations, these include such effects as the baryonic feedback, cloud-in-cloud problem, and the scale-dependent bias \cite{2016arXiv161109787D}. 

However, at higher end, for objects with masses above $10^{15} \, M_\odot$ one does not expect these effects (mergers or baryonic feedback) to be relevant, and so, the high-mass end of the mass function is expected to be accurately recovered with the Simsilun simulation.
For high masses the Simsilun simulation predicts higher number of objects compared to the Tinker mass function. This is also visible at high redshift, which is presented in Fig. \ref{fig2}.

\section{The most massive objects in the Universe}\label{predictions}

Assuming perfect completeness of a survey, the expected number of objects with masses between $M_{{\rm min}} $--$M_{ {\rm max}} $
detected  in a survey of angular coverage ${\rm d} \Omega$ and
redshift range $z_{{\rm min}}$--$z_{{\rm max}}$
can be inferred directly from eq. (\ref{nzm}).
There are two competing effects that contribute to the observed number of objects.
The first one is the volume, i.e. the larger the redshift range the larger the volume, hence larger number of expected objects.
The second effect is related to the fact that the number density of the most massive objects 
decreases with redshift, cf. Fig. \ref{fig2}, where the amplitude of the mass function is significantly lower at higher redshifts. These two competing effects are presented in Fig. \ref{fig3}.

\subsection{Most massive objects at high redshifts}

Figure \ref{fig3} shows the volume covered by the 455 deg$^2$ survey and full-sky survey (as a function of redshift). For comparison, the volume needed to observe 1 object 
with mass $1.77 \pm 0.21 \times 10^{15} M_\odot$ (weighted average of the five most massive ACT clusters), at that redshift is also presented. The upper panel shows the results for the Simsilun mass function and the lower panel for the Tinker mass function.
One should note that this is for comparison only, as these two volumes are not directly related. The volume covered by the survey is the volume within the past lightcone, i.e. this is the volume which is evaluated across different times, from the present-day $t(z=0)$ to a particular redshift $t(z)$.
In contrast, the volume needed to observe 1 object is the volume evaluated at a 
single time instant, i.e. this is the volume evaluated at that particular redshift.
If the number density were constant and did not change with time, these two volumes would be directly comparable; i.e. if the volume covered by the survey were equal to the volume needed to observe 1 object at that redshift, then we would expect to see 1 objects within the survey.  However, since the number density decreases with redshift, thus when the volume covered by the survey is equal to the volume needed to observe 1 object at that redshift then we should expect to see at least 1 object (if not more).
Consequently, if the volume covered by the survey is always smaller than
the volume needed to observe 1 object, then it is unlikely to expect to see any such objects in the survey.

As seen from Fig. \ref{fig3} there are two regimes sensitive to the mass function.
The first regime is the low-redshift universe, where low-volumes prevent observing
most massive objects (since their number density is small one requires large volume to observe these rare objects). The second one is the high-redshift universe where the most massive objects are not sufficiently evolved yet; consequently their very small number density makes them extremely rare, despite large volumes. 

At high redshift the most effective method of detecting clusters is by the observations of the Sunyaev--Zel'dovich effect. 
However, this method does not allow to estimate the mass of the observed cluster, and so one needs to implement other methods to estimate the mass.
For the Atacama Cosmology Telescope (ACT) survey, the mass of the clusters
(detected with the Sunyaev--Zel'dovich effect) were estimated using the scaling relations between the observed velocity dispersion and their mass. The five most massive clusters observed within the ACT survey are
ACT-CL J0102-4915 at $z = 0.8701 \pm 0.0009$ with mass $M_{200c} = 1.68 \pm 0.39 \times 10^{15} M_\odot$; 
ACT-CL J0237-4939 at $z = 0.3344 \pm 0.0007$ with mass $M_{200c} = 2.06 \pm 0.43 \times 10^{15} M_\odot$; 
ACT-CL J0330-5227 at $z =0.4417 \pm 0.0008$ with mass $M_{200c} = 1.77 \pm 0.41 \times 10^{15} M_\odot$; 
ACT-CL J0438-5419 at $z =0.4214 \pm 0.0009 $ with mass $M_{200c} = 2.18 \pm 0.52 \times 10^{15} M_\odot$; and
ACT-CL J0559-5249 at $z =0.6091\pm 0.0014 $ with mass $M_{200c} =1.54 \pm 0.44  \times 10^{15} M_\odot$ \cite{0004-637X-772-1-25}. 
The average uncertainty-weighted mass of these 5 clusters is $1.77 \pm 0.21\times 10^{15} M_\odot$. As seen from Fig. \ref{fig3}, for the Tinker mass function, these 5 most massive clusters are not expected to be seen within a 455 deg$^2$ survey, they are only expected to be observed within the full-sky survey. On the contrary, the Simsilun mass function does not have problems with explaining the existence of these objects; also their observed redshift range is in perfect agreement with predictions of the Simsilun simulation.

This discrepancy between the  number of objects with masses $1.77 \pm 0.21\times 10^{15} M_\odot$, expected to be observed in a 455 deg$^2$ survey  is further presented in Fig. \ref{fig4}. Figure \ref{fig4} shows the number of objects evaluated from eq. (\ref{nzm}) with $M_{{\rm min}} = 1.77 \pm 0.21\times 10^{15} M_\odot$ and $z_{{\rm min}} = z$.
The redshift position as well as the expected numbers are also presented in Fig. \ref{fig4}, where upper panel shows the results for the Simsilun mass function and the lower panel for the Tinker mass function. As seen from Fig. \ref{fig4}, the Simsilun simulation with a slightly higher amplitude at the high-mass end of the mass function (compared to the Tinker mass function) has no problems with explaining the observed number of the most massive clusters observed by the ACT survey. On the other hand the Tinker mass function seems to be at odds with the observed data.

\subsection{Most massive objects at low redshifts}

\begin{figure}
\begin{center}
\includegraphics[scale=0.62]{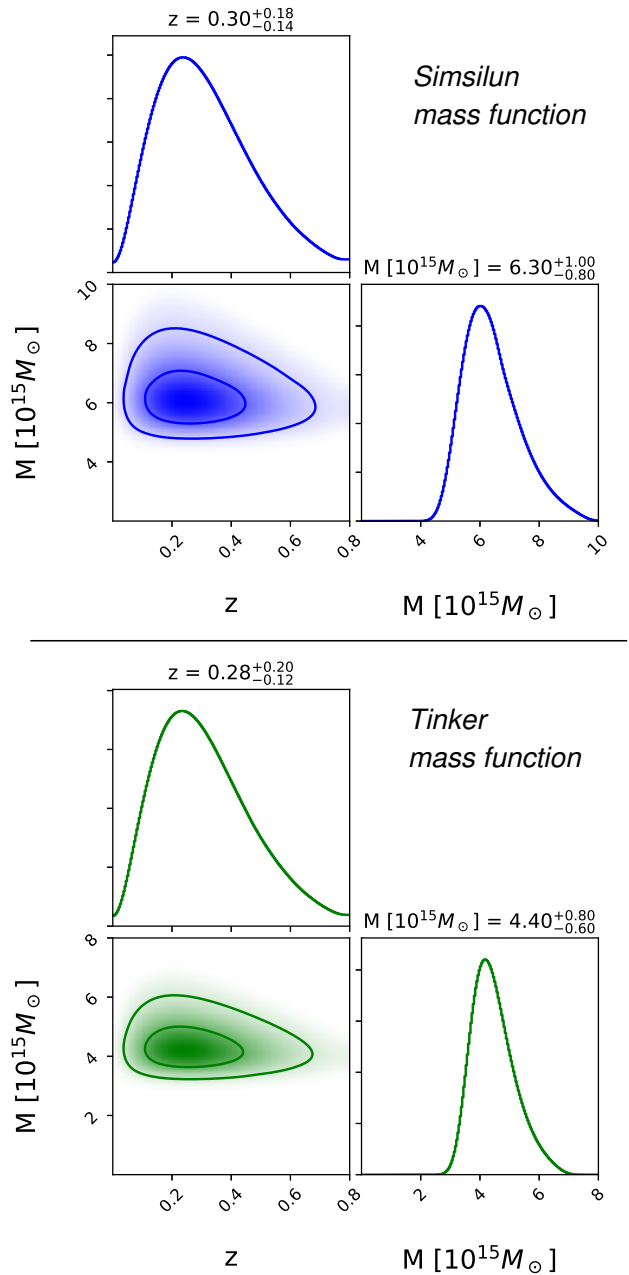}
\end{center}
\caption{Predictions for the most massive object in the Universe obtained using the Simsilun
mass function (upper panels) and Tinker mass function (lower panels).
The density threshold $\Delta$ is set to 200, i.e. the presented predictions are for the quantity $M_{200}$ in units of $10^{15} M_\odot$.
The contours show 1$\sigma$ and 2$\sigma$ regions.}
\label{fig5}
\end{figure}

\begin{figure}
\begin{center}
\includegraphics[scale=0.6]{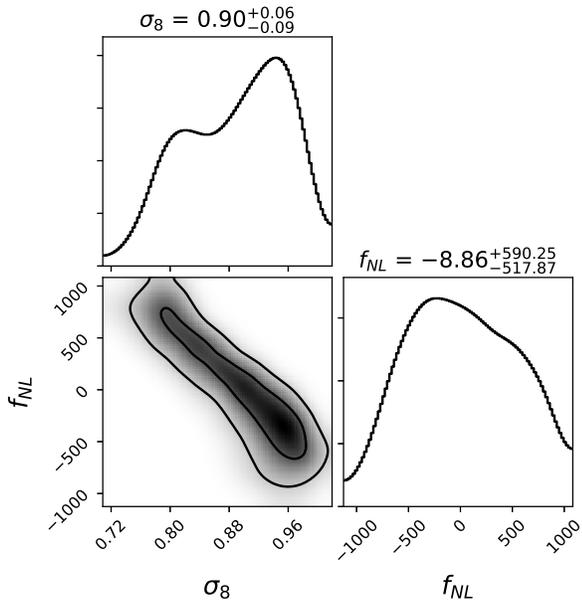}
\end{center}
\caption{The proof-of-concept calculations that show the required change of the parameters $\sigma_8$ and $f_{NL}$ needed to recast the standard mass function obtained within the Newtonian $N$-body simulations onto the mass function from the Simsilun simulation mass function (the Simsilun simulation uses $\sigma_8 = 0.815$ and $f_{NL} = 0$). The contours show 1$\sigma$ and 2$\sigma$ regions. The aim of the figure is to highlight the importance of the relativistic effects related to formation of the most massive objects in the universe -- if these effects are not properly taken into account then they could be mistaken for non-Gausianities (especially if one uses a tight prior on $\sigma_8$).}
\label{fig6}
\end{figure}

Apart from the high redshift, the second regime which is very sensitive to the prediction of the mass function is the number of the most massive objects at low redshifts, where the number of these objects is low due to small volumes. 
The larger the mass, the larger the volume is needed to observe such extreme objects. If the mass is too high, the probability of observing such objects goes to zero, and so there is an upper limit on the mass of the most massive objects. The predictions for the most massive objects obtained using the Simsilun and Tinker mass functions are presented in Fig. \ref{fig5}. 
The most massive object is of the mass $6.3^{+1.0}_{-0.8} \, \times 10^{15} M_\odot$ for the Simsilun simulation, and $4.4^{+0.8}_{-0.6} \, \times 10^{15} M_\odot$ for the Tinker mass function.

If we were to compare these predictions with observations, the most massive supercluster observed up-to-date is most likely the Shapley Concentration,
whose mass is estimated to be between 
$6 \times 10^{15} M_\odot$ \cite{2000MNRAS.312..540B}
and $8 \times 10^{15} M_\odot$ \cite{2006A&A...445..819R},
but likely even higher \cite{2006A&A...447..133P}.
While the existence of such a massive  object is in agreement with the Simsilun simulation, its existence is slightly at odds with the Tinker mass function.

However, if we take into account that the distance to the Shapley Concentration is approximately 200 Mpc, then the probability of observing such a massive object within such a small volume drops. 
Probability of observing an object with mass of at least  $6 \times 10^{15} M_\odot$ within the distance of 200 Mpc is $3 \times 10^{-2}$ and $3 \times 10^{-3}$ for Simsilun and Tinker mass functions respectively. While this could be explained using the extreme value statistics, the probability of observing 
2 very massive objects within the distance of 200 Mpc is very low and poses a challenge for the standard cosmological model \cite{2011MNRAS.417.2938S}.
Indeed, for the Tinker mass function the probability of 
observing 2 objects with masses above $5 \times 10^{15} M_\odot$ and within the distance of 200 Mpc is  $6 \times 10^{-5}$, which makes it inconceivably small. 
For the Simsilun mass function the probability of observing  2 objects with masses above $5 \times 10^{15} M_\odot$ and within the distance of 200 Mpc is  $3 \times 10^{-3}$, which is still within $3\sigma$.
Thus, taking into account the existence of the Great Attractor -- distance between 50 and 80 Mpc and mass  between $4 \times 10^{15} M_\odot$  and $6 \times 10^{15} M_\odot$ \cite{2008GReGr..40.1771B} --
it seems that again (as in the case of the ACT clusters) the predictions 
for the most massive objects obtained from the Tinker mass function are at odds with the data.

\section{Conclusions}\label{conclusions}

This paper investigated the formation of the most massive objects within the framework of the silent universes.
This framework is based on the solution of the Einstein equations obtained within 1+3 comoving coordinates. 

They can be characterized by locally varying expansion rate, density, shear, and curvature. In the limit of spatial homogeneity and isotropy they reduce to the FLRW models. Consequently, local properties differ from the global properties of the universe. For example, overdense regions could be characterized by positive spatial curvature, which could affect the growth rate.

There is a debate, whether this effect could be mitigated by a proper choice of coordinate. It is know that a gauge transformation from comoving coordinates to the Poisson gauge could lead to negligibly small variations in the spatial curvature. 
These transformation are well understood within the linear regime
\cite{2017arXiv171202967B}). However, in the nonlinear regime 
comparison between predictions obtained with the comoving coordinates and Poisson gauge are not trivial (cf. Fig. 1 in Ref. \cite{2013PhRvL.110b1302B}) and in fact such comparisons are questionable (cf. \cite{2010AIPC.1241.1074M}).

If the spatial curvature is a coordinate artifact it should not have any physical impact on the growth of cosmic structure.  
This paper investigated this issue using the Simsilun simulation, which allows for the effect of varying spatial curvature. 
This effect is absent in the standard Newtonian $N$-body simulations where space 
has a uniform spatial curvature throughout the whole simulation. 
The results of our investigations show that indeed we do observe 
a larger number of objects compared to the standard Press-Schechter method, which was implemented also to the result of our simulation. 

We also compared the results to the predictions obtained from using the Tinker mass function.
It should be stressed that such a comparison is less reliable as the Tinker mass function is inferred from the standard Newtonian $N$-body simulations.
However, it is interesting to note that the prediction for the number of the most massive objects obtained using the Tinker mass function is at odds with the observational data, where as the Simsilun simulation correctly predicts the number of most massive ACT clusters, as well as, the existence of the Shapley Concentration and the Great Attractor. 

However, the Simsilun simulation does not allow for rotation or transfer of energy between various regions of the universe, and thus, at sub Mpc-scales where these effects are important the Simsilun simulation is not expected to work well. Therefore, more work is needed especially using more advanced relativistic schemes, such the one based on the fully relativistic Lagrangian framework \cite{2012PhRvD..86b3520B,2013PhRvD..87l3503B,2015PhRvD..92b3512A,2017PhRvD..96l3538A}; the first application
    of this framework to observationally realistic,
    standard $N$-body cosmological initial conditions
    was recently published by \cite{Roukema17silvir}.

Other cosmological relativistic simulations discussed in the literature are 
either based on the weak-field limit \cite{2016JCAP...07..053A}
or  on the Einstein toolkit \cite{2012CQGra..29k5001L}, which implements the BSSN formalism \cite{Bentivegna:2015flc,Mertens:2015ttp,2017PhRvD..95f4028M}. However, most of the studies that implement the BSSN approach use very simple density fields \cite{Bentivegna:2015flc,2017PhRvD..95f4028M,2018PhRvD..97d3509E} and do not focus on the structure formation and the evolution of the most massive cosmological objects, which is not a trivial task \cite{2001ApJ...558L..79K,2017mgm..conf.2333O}.

The results of this paper highlight the importance of the non-linear, relativistic effects on structure formation and emphasize that the understanding of these effects is vital if we want to use the most massive objects as a cosmological probe.
As a proof-of-concept, to show how important these effects can be, we take the Simsilun mass function and fit a mass function of 
\cite{2010MNRAS.402..191P} which was obtained from the Newtonian $N$-body simulations with 
 non-Gaussian initial conditions.  We keep all cosmological parameters fixed except for the parameters 
$\sigma_8$ and $f_{NL}$, and we fit the \cite{2010MNRAS.402..191P} mass function. The mass range over which we perform the fit is 
 $10^{15} - 10^{16} M_\odot$. The result of this procedure is presented in Fig. \ref{fig6}.
The results presented in Fig. \ref{fig6} show what kind of change of cosmological parameters is required 
in order to reconcile the mass function obtained from Newtonian simulations with the semi-relativistic Simsilun simulation.
In other words, suppose that the Simsilun simulation correctly captures the properties of the universe on its largest scales and the evolution of most massive objects. Then, in such a case, if one fits the mass function obtained using the Newtonian simulations to observations (here the Simsilun simulation) then such a procedure  biases the cosmological parameters. For example in order to explain the observed number of high-redshift massive clusters  (cf. Fig. \ref{fig3}) or the existence of the most massive objects (cf. Fig. \ref{fig5}),
one would either overestimate the parameter $\sigma_8$  or (if the parameter  $\sigma_8$ is fixed with the CMB prior) one would require non-Gaussian initial conditions. 
It should be noted, that in the presented example we do not vary other cosmological parameters or combine them with other cosmological probes, such as for example the CMB observations. The results presented in Fig. \ref{fig6}
are the proof-of-principle calculations to show that if one does not take these effects into account then one could bias the  cosmological parameters.

In summary, the results of this paper suggest that the nonlinear relativistic effects could affect the formation of the most massive cosmological objects, leading to a relativistic environment-dependence of the growth rate of the most massive clusters. Our study was based on a simplified model, therefore more investigation of this phenomenon is needed. If indeed the 
 environment-dependence of the growth rate of the most massive clusters is confirmed by more detailed study, these these effects will be of 
significance  when analyzing the data from the next-generation satellite missions such as Euclid \cite{2013LRR....16....6A} and eROSITA \cite{2012arXiv1209.3114M}.

\section*{Acknowledgements}

KB acknowledges the support of the Australian Research Council through the Future Fellowship FT140101270. JJO is supported by Lyon Institute of Origins (LIO) under Grant No. ANR-10-LABX-66 and by  National Science Centre, Poland, under grant 2014/13/B/ST9/00845.
This work is part of a project that has received funding from the European Research Council (ERC) under the European Union's Horizon 2020 research and innovation programme (grant agreement ERC advanced grant 740021-ARTHUS, PI: T. Buchert).  Computational resources used in this work were provided by the ARC (via FT140101270) and the University of Sydney HPC service (Artemis). The corner plots were created using the software \texttt{Corner} \cite{corner}.
 Finally, discussions and comments from Thomas Buchert, Boud Roukema, and David Wiltshire are gratefully acknowledged.

\bibliography{funm} 
\bibliographystyle{apsrev}

\end{document}